# Nucleation of liquid droplets in a fluid with competing interactions

A J Archer[a] and R Evans[b]

[a]*Department of Mathematical Sciences, Loughborough University, Loughborough LE11 3TU, UK*

[b]*H H Wills Physics Laboratory, University of Bristol, Bristol BS8 1TL, UK*

**Abstract:** Using a simple density functional theory (DFT) we determine the height of the free energy barrier for forming a droplet of the liquid phase from the metastable gas phase for a model colloidal fluid exhibiting competing interactions. The pair potential has a hard core of diameter $\sigma$, is attractive Yukawa at intermediate separations, and is repulsive Yukawa at large separations. We find that even a very weak long-range repulsive tail in the pair potential has a profound effect on nucleation: increasing the amplitude of the repulsive Yukawa tail reduces significantly the free energy barrier height and therefore increases the liquid droplet nucleation rate. The method we introduce for calculating the droplet density profile and free energy employs a fictitious external potential to stabilize a liquid droplet of the desired size, i.e. with a given excess number of particles. For the critical droplet, corresponding to an extremum of the grand potential, this fictitious potential is everywhere zero. We examine the decay of the droplet density profiles into the bulk gas. For a range of nucleation state points the DFT predicts exponentially damped, long wavelength oscillatory decay for systems exhibiting long-range repulsion, contrasting sharply with the monotonic decay found when the pair potential has only an attractive Yukawa piece. The changes in nucleation properties that we find for small amplitudes of the repulsive Yukawa tail reflect the propensity of the fluid to form modulated structures such as clusters or stripes.

1. **Introduction.**

In 2000 D Pini, Ge Jialin, A Parola and L Reatto [1] investigated the equilibrium properties of a model fluid described by the pair potential

$$\beta v(r) = \infty \qquad\qquad\qquad\qquad\qquad\qquad\qquad\qquad r \leq \sigma$$
$$= -\varepsilon\sigma\exp(-Z_1(r/\sigma-1))/r + A\sigma\exp(-Z_2(r/\sigma-1))/r \quad r > \sigma \qquad (1)$$

where $\beta=1/k_BT$ is the inverse temperature, $\sigma$ is the hard sphere diameter and the dimensionless amplitudes $\varepsilon$ and $A$ are both positive. The range parameters are chosen such that $Z_2 < Z_1$ so that the repulsive Yukawa tail decays more slowly than the attractive Yukawa contribution. For a fixed value of $A$ the parameter $\varepsilon$, that governs the strength of the attractive contribution, plays a role somewhat akin to an inverse temperature. A similar model incorporating competing attractive and repulsive interactions but with exponential rather than Yukawa contributions had been introduced earlier [2] to describe two-dimensional systems of certain metallic colloidal particles coated by surfactant that exhibit globular cluster and stripe phases. Ref [1] was the first of several important papers by Luciano Reatto and co-workers [1, 3-8] on the behaviour of model colloidal particles interacting via mermaid potentials (an attractive head and a repulsive tail). Such potentials are germane to charged colloids dispersed in a solvent containing non-adsorbing polymer: short-ranged attraction arises from the depletion of the polymer while the screened Coulomb repulsion is sufficiently strong that the ultimate decay of the effective interaction potential is a repulsive Yukawa. Mermaid potentials can also arise in other soft-matter systems; for an overview and references to early theoretical treatments see Archer and Wilding [9]. A key feature of such models is that they can undergo transitions to stable inhomogeneous phases, e.g. cluster or stripes, when $A$, the strength of the repulsive tail, is sufficiently large. Recall that for $A=0$ the model reduces to the much-studied hard-core, one-Yukawa fluid which, provided the attractive potential is not too 'sticky', exhibits the conventional phase behaviour of a simple fluid, namely solid, liquid and gas phases.

In the present study we consider the same class of fluid as that investigated in [1], i.e. we study the model where the amplitude $A$ is sufficiently small that the fluid still exhibits a stable gas-liquid transition but is sufficiently large that the fluid approaches the stability limit of this transition – for slightly larger values of $A$ the gas-liquid transition will be replaced (at sufficiently high $\varepsilon^{-1}$) by a transition to an inhomogeneous phase [1, 6-9]. Pini *et al*. [1] studied the thermodynamic properties and the static structure of the fluid above the critical point and found that these are very different from what is observed in a simple fluid in the corresponding region of the phase diagram. The main results from [1], where the authors set $Z_1$=1 and $Z_2$=0.5, are (i) the size of the region in the temperature-density ($T$-$\rho$) plane where the compressibility is very large is much bigger than the near-critical region of a simple fluid, (ii) the top part of the liquid-gas coexistence curve becomes very flat and (iii) the pair correlation function $g(r)$ in the region of large compressibility has faster decay at long distance than in the case of one-Yukawa attraction but is strongly enhanced at short and intermediate distance – out to several tens of $\sigma$. As pointed out in Ref [6] these features can be regarded as precursors of the inhomogeneous phases that occur for larger values of $A$. Using the SCOZA (self-consistent Ornstein Zernike approximation) and HRT (hierarchical reference theory) integral equation approaches Pini *et al*. [1] estimate the liquid-gas transition of the model (1) becomes unstable for $A/\varepsilon \approx 0.097$ when $Z_1$=1 and $Z_2$=0.5.(Note that the convention we use for the parameter $A$ in (1) is the same as in [7,8,9] while in [1,6] the amplitude of the repulsive contribution is $A\varepsilon$.)

Whereas Ref [1] focussed primarily on liquid-gas coexistence and supercritical states here we focus on subcritical states and consider the (homogeneous) nucleation of a liquid droplet from the bulk gas phase. We are concerned mainly with the issue of how the nucleation free energy barrier $\Delta\Omega$ for droplet formation varies with the parameter $A$, i.e. we seek to understand how critical droplet formation depends on the strength of the repulsive Yukawa tail. The approach we take is based on a simple density functional theory (DFT) that is in the same spirit as the early study by Oxtoby and Evans [10] on liquid-gas nucleation for the hard-core, one-Yukawa fluid.

Our paper is arranged as follows: in Section 2 we give some brief background to classical nucleation theory and provide some details of the DFT that we employ. We also describe how the DFT calculations were implemented to obtain droplet density profiles and excess grand potentials as a function of $N_{ex}$, the excess number of particles in the droplet. Our method, which employs a fictitious external potential to stabilize a droplet of a given size, is different from that used by other authors in recent studies and we discuss its advantages and limitations. Section 3 gives results for droplet profiles and for free energy barriers $\Delta\Omega$ as a function of supersaturation for different values of $A$. We find that the strength of the repulsion has a profound effect on the height of the nucleation barrier. In Section 4 we make some concluding remarks.

## 2. Theory

### 2.1. Classical nucleation theory: a reminder

In the classical treatment of nucleation one considers a reservoir of gas that is supersaturated, i.e. it is at a fixed chemical potential $\mu > \mu_{bin}(T)$, the value at the binodal where the bulk gas and liquid coexist. The corresponding density is $\rho_b(\mu)$ and the pressure is $p_b(\mu)$. If a droplet of liquid with radius $R$ and pressure $p_l(\mu)$-the pressure of the liquid at the same chemical potential- forms the grand potential of the system is approximated by

$$\Omega = -p_l(\mu) 4\pi R^3 / 3 - p_b(\mu)(V - 4\pi R^3 / 3) + \gamma 4\pi R^2 \qquad (2)$$

where $\gamma$ is the (planar) liquid gas surface tension and $V$ is the total volume. Since the grand potential of the uniform gas phase is $-p_b(\mu)V$ the excess grand potential for droplet formation is simply

$$\Delta\Omega = -\Delta p \, 4\pi R^3 / 3 + \gamma 4\pi R^2 \qquad (3)$$

where $\Delta p \equiv p_\ell(\mu) - p_b(\mu) > 0$. The critical droplet forming the barrier to nucleation corresponds to the maximum of $\Delta\Omega$; its radius $R_c$ is given by

$$\Delta p = 2\gamma / R_c \qquad (4)$$

and the barrier height is

$$\Delta\Omega(R_c) = \frac{16\pi}{3} \gamma^3 / (\Delta p)^2 \qquad (5)$$

Note that Eq (4) is the Laplace condition for the mechanical stability of the droplet and it is straightforward to show [10] that $\Delta\Omega = \Delta G$, the excess Gibbs free energy appropriate to nucleation at fixed pressure. As the rate of formation of critical droplets is assumed to be proportional to exp $[-\beta \Delta\Omega (R_c)]$, equation (5) determines nucleation rates for given $\mu$, $T$.

### 2.2. DFT approach

Equation (2) is, of course, based on macroscopic (capillarity) ideas and it is difficult to justify their use for microscopic droplets; a typical critical droplet has an excess number of particles (atoms) lying in a range of a few tens to a few hundred. DFT provides a natural *microscopic* framework for tackling the calculation of nucleation barriers. Specifically one writes the grand potential of the fluid as a functional [11] of the inhomogeneous density $\rho(\mathbf{r})$:

$$\Omega[\rho] = F[\rho] - \mu \int d\mathbf{r} \rho(\mathbf{r}) + \int d\mathbf{r} V_{ext}(\mathbf{r}) \rho(\mathbf{r}) \qquad (6)$$

where $F[\rho]$ is the intrinsic Helmholtz free energy functional and $V_{ext}(\mathbf{r})$ denotes any external potential that is present.

In the DFT approach the critical droplet has a density profile $\rho_c(\mathbf{r})$ that corresponds to the solution of

$$0 = \left.\frac{\delta \Omega[\rho]}{\delta \rho(\mathbf{r})}\right|_{\rho_c} = \left.\frac{\delta F[\rho]}{\delta \rho(\mathbf{r})}\right|_{\rho_c} - \mu \qquad (7)$$

i.e. an extremum of $\Omega[\rho]$ with $V_{ext}(\mathbf{r}) \equiv 0$. The profile is taken to be spherical $\rho(\mathbf{r}) \equiv \rho(r)$ and the goal is to find a solution of (7) that satisfies the boundary condition:

$$\lim_{r \to \infty} n(r) = 0 \qquad (8)$$

where

$$n(r) \equiv \rho(r) - \rho_b(\mu) . \qquad (9)$$

Unlike in the majority of applications of DFT to interfacial phenomena, where the external potential stabilizes the interface and one seeks the minimum of (6), in nucleation the solution of (7) corresponds to a saddle point in function space [10]. This leads to some subtleties in finding solutions and in the interpretation of non-critical droplet profiles. Before describing the approach that we have implemented we mention some earlier approaches to the problem.

In their original DFT study Oxtoby and Evans [10] employed a trial and error iteration process that begins with an initial guess for the profile

$$\begin{aligned}\rho(r) &= \rho_l & r > R \\ &= \rho_b & r \geq R\end{aligned} \qquad (10)$$

and iterates the Euler-Lagrange equation (7). If the guessed radius $R$ is too small the droplet shrinks and the profile evolves to a uniform gas of constant density $\rho_b(\mu)$. On the other hand if the guess for $R$ is too large the droplet grows indefinitely and evolves to the (stable) uniform liquid of density $\rho_\ell(\mu)$. Oxtoby and Evans used this property, starting with various initial guesses for $R$ and iterating a (small) number of times to find $\Delta\Omega$ as a function of $R$ and then determined its maximum, essentially by inspection. Although this scheme will in principle identify the correct critical droplet it is not particularly robust; the number of iterations is somewhat arbitrary and one can miss the saddle point if one iterates too many or too few times.

In a subsequent DFT study Talanquer and Oxtoby [12] introduced a method that allowed them to study non-critical droplets. They extremize $\Omega[\rho]$ with the constraint that the total number of particles in a sub-volume is fixed at a given number. They also adjusted the density outside this volume so that the profile is continuous. Lutsko [13, 14] has argued that this method of solution 'amounts to changing the chemical potential so as to stabilize a cluster with the prescribed number of atoms' and that the resulting clusters are not equivalent to clusters of different sizes at the given chemical potential $\mu$. Other authors, e.g. Uline and Corti [15], have introduced similar constraints but that do not adjust the density outside the droplet volume. However, as pointed out by Lutsko [13, 14], these have their own disadvantages.

It is instructive to note that the simplest constraint method one might choose runs into difficulties. Suppose we constrain the excess number of particles in the droplet to be fixed at $N_{ex}$, i.e. we require

$$\int d\mathbf{r}\, n(r) = 4\pi \int dr\, r^2 n(r) = N_{ex} \qquad (11)$$

where the integral is over the total system volume $V$. We would then extremize the functional

$$F[\rho] - \mu \int d\mathbf{r}\, \rho(r) - \alpha \left[ \int d\mathbf{r}\, n(r) - N_{ex} \right]$$

which leads to the Euler-Lagrange equation

$$0 = \frac{\delta F[\rho]}{\delta \rho(r)} - (\mu + \alpha), \qquad (12)$$

i.e. the chemical potential shifted by a constant amount $\alpha$, the Lagrange multiplier. If $\alpha \neq 0$ then clearly (8) cannot be satisfied [14]. Only in the special case of the critical droplet will (8) be satisfied and then $\alpha = 0$.

Lutsko [14] proposed a new 'structural' constraint that defines the size of a droplet (he actually considered a bubble of gas) as the radius at which the local density becomes less than some value $\rho^* \approx (\rho_\ell + \rho_g)/2$ and constrains the volume $\Gamma$ of the droplet. He shows the method works for all $\Gamma$ but yields discontinuous density profiles. Only for the critical droplet is the resulting profile continuous and this corresponds to a Lagrange multiplier equal to zero. Ref [14] compares DFT results for $\Delta\Omega$ as a function of $N_{ex}$ obtained from this structural constraint with those obtained by i) parameterizing the profile (and minimising w.r.t. the parameters) and ii) implementing the so-called nudged elastic band (NEB) method which is a powerful chain-of-states method often used to determine chemical reaction paths. The NEB should provide an unbiased and robust determination of the minimum free energy path (MFEP) between metastable states. Lutsko [14] finds that $\Delta\Omega$ is generally higher from his structural constraint method than from NEB. However, all three methods yield the same barrier height and value of $N_{ex}$ for the critical bubble. Lutsko considers a Lennard-Jones fluid with $k_B T/\varepsilon_{LJ} = 0.8$.

Here we implement a different method that is closely related to that used by Hopkins *et al.* [16] in recent DFT studies of the van Hove function where normalization constraints are placed on the 'self' density profile $\rho_s(r,t)$ and the 'distinct' profile $\rho_d(r,t)$ with $t$ denoting time. The idea is to find an effective potential $V_{eff}(r)$ that self-consistently, and on the fly as part of an iterative numerical procedure, determines solutions of

$$0 = \frac{\delta F[\rho]}{\delta \rho(r)} - \mu + V_{eff}(r) \tag{13}$$

*and* satisfies (8) and (11). Clearly we require

$$\lim_{r \to \infty} V_{eff}(r) = 0 \tag{14}$$

Otherwise $V_{eff}(r)$ is a priori unknown. Only for the critical droplet, with the 'exact' critical value of $N_{ex}$, is $V_{eff}(r)$ identically zero.

The implementation begins by re-writing (13) as

$$\rho(r)/\rho_b = \exp[c^{(1)}(r) - c^{(1)}(\infty) - \beta V_{eff}(r)] \qquad (15)$$

with the one-body direct correlation function

$$c^{(1)}(\mathbf{r}) \equiv c^{(1)}[\rho;\mathbf{r}] = -\beta \frac{\delta F^{ex}[\rho]}{\delta \rho(\mathbf{r})} \qquad (16)$$

where

$$F^{ex}[\rho] = F[\rho] - F^{id}[\rho] \qquad (17)$$

is the excess, over ideal gas, free energy functional. We choose a step-function profile (10) with $R$ chosen to give the required $N_{ex}^{req}$ through (9) and (11) as the initial profile in an iterative procedure. At each step we insert the profile from the previous iteration $n^{old}(r)$ into the right hand side of (15), with $V_{eff}(r) = 0$, to generate a new (continuous) profile

$$n^{new}(r)/\rho_b = \exp[c^{(1)}[n^{old} + \rho_b ;r] - c^{(1)}(\infty)] - 1$$

which corresponds to a value $N_{ex}^{new} \neq N_{ex}^{req}$. The new profile is rescaled by replacing $n^{new}(r)$ by $n^{new}(r) N_{ex}^{req} / N_{ex}^{new}$ and iteration proceeds by mixing old and new solutions, i.e. the profile is replaced by $\lambda$ x *new* + (1 - $\lambda$) x *old* where the mixing parameter $\lambda$ varies between 0.005 and 0.1, with small values used for the first few iterations – a standard Picard procedure. One repeats the steps until a satisfactory convergence criterion is met. Note that it is the *difference* $n(r)$, see (9), that is rescaled so that (8) is always satisfied; one is not imposing an additional chemical potential $\alpha$ as was the case with the Lagrange multiplier method in (12).

Rather one can enquire at the end of the iteration what the self-consistent potential is that generated the converged density profile by substituting the latter into (15). For a given $N_{ex}^{req}$, $V_{eff}(r)$ will, in general, be non-zero and one can interpret this function as the effective external potential required to stabilize a non-critical droplet. For values of $N_{ex}^{req}$ that lie very close to that of the critical droplet the resulting $V_{eff}(r)$ should be small for all values of $r$. Indeed this is what we find in our calculations for a particular approximate DFT – see Section 3.

One can ask whether for a given $N_{ex}$ (and a given DFT) a non-critical droplet profile generated by the present method has a lower excess grand potential $\Delta\Omega$ than is obtained using other constraints such as that of Lutsko [14]. We have not made such comparisons. Nor have we compared our results with those from the NEB method since our main focus is on determining the barrier height for the critical droplet. Such systematic comparisons may shed valuable insight into how well our method accounts for the MFEP.

**2.3. A simple mean-field DFT for the mermaid potential (1)**

The previous sub-section outlined the general strategy for tackling nucleation within the context of DFT. Here we describe the simple DFT that we used to perform calculations for the model potential (1).

We approximate the excess Helmholtz free energy functional by

$$F^{ex}[\rho] = F_{hs}^{ex}[\rho] + \frac{1}{2}\iint d\mathbf{r}\, d\mathbf{r}'\rho(\mathbf{r})\rho(\mathbf{r}')v_p(|\mathbf{r}-\mathbf{r}'|) \tag{18}$$

where the first term refers to the functional for a reference fluid of hard-spheres with diameter $\sigma$ while the second term treats the remaining, non-hard core part of the potential within the simplest mean-field approximation. The 'perturbation' potential is defined as

$$\beta v_p(r) = \begin{cases} -\varepsilon + A & r \leq \sigma \\ \beta v(r) & r > \sigma \end{cases} \tag{19}$$

Note that $-\varepsilon + A$ is the value of $\beta v(r)$ at contact, $r = \sigma^+$. The functional (18) was employed in earlier studies of the bulk and interfacial structure of the model fluid [7]. Taking two functional derivatives of (18) generates the random-phase approximation (RPA) for the pair direct correlation function of the uniform fluid:

$$c^{(2)}(r) = c_{hs}^{(2)}(r) - \beta v_p(r) \tag{20}$$

When $F_{hs}^{ex}[\rho]$ is the non-local, weighted-density Rosenfeld fundamental measure hard-sphere free energy functional the resulting $c_{hs}^{(2)}(r)$ is the Percus-Yevick pair direct correlation function [17].

In our present study we make a further approximation in order to simplify the numerics. We employ a local density approximation for $F_{hs}^{ex}[\rho]$:

$$F_{hs}^{ex}[\rho] \approx \int d\mathbf{r}\rho(\mathbf{r}) f_{hs}^{ex}(\rho(\mathbf{r})), \qquad (21)$$

where $f_{hs}^{ex}(\rho)$ is the excess over ideal Helmholtz free energy per particle of the uniform hard-sphere fluid with density $\rho$. Such an approximation is not able to describe the oscillatory structure of density profiles that arise for liquids near confining walls where the oscillations due to packing of the particles have a wavelength $\sim\sigma$; one requires the full non-local functional to account for these.

However, as demonstrated in [8], if one is concerned with inhomogeneities that occur on significantly larger length scales, such as in stripe and cluster phases, the local density approximation captures the key features of the structure and thermodynamics. Note that, as in [8], we employ the accurate Carnahan-Starling approximation for the hard-sphere free energy:

$$\beta f_{hs}^{ex}(\rho) = \eta(4-\eta)/(1-\eta)^2 \qquad (22)$$

where $\eta = \pi\rho\sigma^3/6$ is the packing fraction, rather than the Percus-Yevick compressibility equation state that results from use of (18) with the Rosenfeld functional. We should recognize that employing (21) in (18) does incorporate any oscillatory structure in density profiles that is set by the form of the perturbation potential $\beta v(r)$ – at least at mean-field level.

Regarding the problem of liquid nucleation our approach neglects any oscillatory structure in the profiles of droplets occurring on the length scale $\sigma$ but does capture longer wavelength oscillations that might be present. Recall that the simplest square-gradient [11] description of nucleation does not capture any oscillatory structure. In the next section we show that the tails of droplet profiles do exhibit long-wavelength oscillations when the fluid is close to the onset of an inhomogeneous phase.

Finally we remark that the present DFT with the local density approximation (21) is the same as that used by Oxtoby and Evans [10] for the one-Yukawa, $A=0$, case. However, in [10] $v_p(r)$ was not truncated inside the hard-core, i.e. the Yukawa potential was extended down to $r=0$.

3. **Results of DFT Calculations**

The procedures described in the previous section were carried out for the model fluid (1). We set $\beta = 1$ in all the calculations. In Fig 1 we plot phase diagrams in the $\varepsilon^{-1}$, $\rho_b$ plane for $Z_1 = 1$ and $Z_2 = 0.5$. The bulk free energy is determined from (18, 19 and 21) from which we determine the liquid-gas coexistence curves. Consistent with the results of [1] the bulk critical value $\varepsilon_c^{-1}$ decreases as the amplitude of the repulsion, $A$, increases. Within this RPA DFT treatment the bulk critical density $\rho_c\sigma^3 = 0.24913$ is independent of $A$ and it is easy to show that having calculated the binodal and spinodal for one value of $A$, results for all other values can be obtained by a simple rescaling of the vertical ($\varepsilon^{-1}$) axis [7]. A detailed comparison between RPA DFT results and those from SCOZA was given in Fig 4 of [7] for the case $A = 0.02$; quite good agreement between the two theories was found regarding the locations of the binodal, spinodal and the two Kirkwood lines. Of course this DFT is a mean-field theory so it is unable to describe the correct shape of the coexistence curve in the critical region.

We restrict $A$ to values $\leq 0.06$. For larger values the RPA DFT exhibits a $\lambda$ line at which the uniform fluid becomes unstable w.r.t. periodic density fluctuations [7, 8].The $\lambda$ line, when it occurs, takes the shape of a loop that crosses both branches of the binodal and then meets and is bounded by the spinodals. The case $A = 0.06$ does in fact exhibit a $\lambda$ line. However, on the scale of Fig 1 this line is indistinguishable from the spinodal: the maximum of the $\lambda$ line lies at $\varepsilon^{-1} = 1.8480$ whereas the critical point is at $\varepsilon_c^{-1} = 1.8468$. Thus the value $A = 0.06$ is essentially the threshold where the $\lambda$ line first appears in the present theory. The phase diagram is very similar to that in Fig 5 of [7] but the $\lambda$ line is much closer to the spinodal.

In our nucleation calculations we consider two 'temperatures' $\varepsilon^{-1} = 0.8\,\varepsilon_c^{-1}$ and $0.7\,\varepsilon_c^{-1}$ and investigate droplets and their excess grand potential as a function of $N_{ex}$ for different values of the parameter $A$. In Fig 2a we plot the density profiles for $A=0$, the one-Yukawa fluid at $\varepsilon^{-1} = 0.8\,\varepsilon_c^{-1}$ and fixed bulk gas density $\rho_b\sigma^3 = 0.09$, a state point roughly mid-way between the binodal and spinodal – see Fig 1. The profiles for $N_{ex} = 360$, 400 and 500 have shapes characteristic of droplets whereas for $N_{ex} = 340$ the profile is almost flat, corresponding to spreading the excess number of particles throughout the system. $N_{ex} = 400$ corresponds roughly to the critical droplet. In Fig 2b we plot the same data but as $rn(r) \equiv r(\rho(r) - \rho_b)$ on a logarithmic scale in order to illustrate the exponential decay of the tails of the profiles. One can observe that the decay length of the profiles is different for different values of $N_{ex}$. In Fig 3a we show the effective potentials $V_{eff}(r)$ that are required to generate these density profiles, as explained in Section 2.2. For the near-critical droplet $N_{ex} = 400$, $\beta V_{eff}(r) < 0.004$ for all $r$ whereas for the largest droplet $N_{ex} = 500$, $\beta V_{eff}(r)$ is large and positive for $r \leq 10\sigma$, i.e. in the interior region of the droplet. For $N_{ex} = 360$, a droplet that is smaller than critical, $\beta V_{eff}(r)$ is large and negative in the interior. Fig 3b displays the results on a logarithmic scale. As with the profiles there is exponential decay with different decay lengths for different $N_{ex}$ – except for $N_{ex} = 340$ where $\beta V_{eff}(r)$ decays very slowly throughout the system volume.

The density profiles for $A=0.06$ at $\varepsilon^{-1} = 0.8\, \varepsilon_c^{-1}$ and fixed bulk gas density $\rho_b \sigma^3 = 0.06$ are shown in Fig 4a for five values of $N_{ex}$. The critical droplet for this state point, which is at a location in the phase diagram similar to that considered for $A = 0$, is close to $N_{ex} = 140$. Note that the profile for $N_{ex} = 80$ is almost flat. Strikingly and in sharp contrast to the droplets with $A=0$, Fig 4b shows that the tails of the density profiles for the larger values of $N_{ex}$ exhibit exponentially damped oscillatory behaviour. The wavelength of the oscillations is about $32\sigma$ for $N_{ex} = 140$. The corresponding effective potentials are plotted in Fig 5a. As previously the large droplets exhibit a repulsive potential in their interior while the sub-critical droplet with $N_{ex} = 100$ exhibits an attractive potential. For $N_{ex} = 140$, $|\beta V_{eff}(r)| < 0.001$ for all $r$ and has decayed to $10^{-6}$ by $r = 20\sigma$ confirming that this droplet is indeed close to critical. The wavelength of the oscillations in $\beta V_{eff}(r)$ is the same as that of $rn(r)$.

The presence of (weak) long wavelength oscillations in the tails of the profiles of 'free' liquid droplets is, at first sight, surprising. We emphasize that these oscillations reflect the propensity of the fluid to form modulated structures such as clusters or stripes. Recall that for sufficiently large $A$ inhomogeneous phases develop. In order to understand fully the origin of the oscillatory behaviour we enquire about the asymptotic decay of the bulk pair correlation function $g(r)$ in different regions of the phase diagram. This topic was discussed in [7] where we found that for mermaid potentials with small but non-zero values of the amplitude $A$, there are *two* Kirkwood [18] lines in the phase diagram at which the asymptotic decay of $g(r)$ crosses over from monotonic to long wavelength oscillatory. An example is given in Fig 4 of Ref [7] for $A=0.02$. The upper Kirkwood line (1) has a maximum far above $\varepsilon_c^{-1}$ and crosses both branches of the binodal. The lower Kirkwood line (2) lies just above the spinodal and its maximum lies slightly above the critical point. Both lines are bounded by the spinodal. As *A* increases Kirkwood line (2) becomes even closer to the spinodal. In the region of the phase diagram between Kirkwood lines (1) and (2) the decay of $r(g(r)-1)$ is exponentially damped oscillatory. Above Kirkwood (1) and in the very narrow region between Kirkwood (2) and the spinodal the decay is monotonic exponential. This means that there is only a very small supercritical region where the asymptotic decay of $r(g(r)-1)$ is exponential, i.e. Ornstein Zernike like.

In the RPA DFT treatment there is a region of state points lying between the gaseous binodal and spinodal that also lies between the two Kirkwood lines. For such (metastable) states we expect the pair correlation function to decay in a damped oscillatory fashion. Note that such oscillatory states are also found within the sophisticated SCOZA theory for $A=0.02$. Moreover the Kirkwood lines obtained from the two theories are rather close [7]. The nucleation state point $\varepsilon^{-1}= 0.8\varepsilon_c^{-1}$, $\rho_b\sigma^3 = 0.06$ that we consider here falls within the oscillatory region between the Kirkwood lines for a range of values of *A* and we confirmed that for $A= 0.04$ the droplet density profile decays in a damped oscillatory fashion.

For $A=0.06$, the case we analyse in detail, there is only one (upper) Kirkwood line. The lower one is replaced by the $\lambda$ line [7], which for this value of *A* lies almost on top of the spinodal, and in the region between these two lines the decay of $g(r)$ is long wavelength oscillatory. The nucleation state point lies in this region.

From general arguments about the asymptotic decay of one-body equilibrium density profiles and pair correlation functions [19] we would expect the density profile of the liquid droplet to decay to the bulk gas density in the same fashion as $g(r)$, i.e. as:

$$n(r) \equiv \rho(r) - \rho_b \sim \tilde{B}\exp(-\tilde{\alpha}_0 r)/r, \qquad r \to \infty \qquad (23)$$

in a region of monotonic decay and as

$$n(r) \sim B\exp(-\alpha_0 r)\cos(\alpha_1 r - \theta)/r, \qquad r \to \infty \qquad (24)$$

in a region of oscillatory decay – assuming of course these arguments hold for metastable states! The inverse decay length $\tilde{\alpha}_0$ corresponds to a pure imaginary pole $i\tilde{\alpha}_0$ of the Fourier transform of $g(r)-1$ at the given bulk state point whereas $\alpha_0$ and $\alpha_1$ correspond to the complex poles $i\alpha_0 \pm \alpha_1$ with the smallest value of $\alpha_0$ - see Ref [7].

Equations (23) and (24) pertain to a 'free' droplet that is subject to a vanishingly small external potential. Thus we might conjecture that within a DFT context they pertain to the critical droplet for which $V_{eff}(r) = 0$, but not to the non-critical droplets. We tested this conjecture as follows: the relevant poles were calculated from the present DFT using the methods of [7, 8] yielding values $\alpha_0\sigma =0.4653$ and $\alpha_1\sigma = 0.1955$, corresponding to a wavelength of $32.1\sigma$, at the nucleation state point. These values were then inserted into the asymptotic form (24). Adjusting the amplitude $B$ and phase $\theta$ provided an excellent fit to the results for the near-critical droplet profile with $N_{ex} =140$ for $r>5\sigma$, i.e. the period of the oscillations and the decay length are indeed determined by the pole analysis. One can see from Fig 4b that the decay lengths $\alpha_0^{-1}$ are different for the other (non-critical) values of $N_{ex}$.

We turn now to results for the excess grand potential $\Delta\Omega$ of the droplets. This quantity is plotted in Fig 6 for $A=0$ and $A=0.06$ at the state points corresponding to the density profiles in Figs 2 and 4. In the one-Yukawa case, $A=0$, one can see that the maximum of $\Delta\Omega$ versus $N_{ex}$ lies close to $N_{ex} = 400$ whereas for $A=0.06$ the maximum is close to $N_{ex} = 140$. As we identify the critical droplet with the maximum it is pleasing that these values of $N_{ex}$ do correspond to droplets having very small effective potentials $V_{eff}(r)$. Note that for small values of $N_{ex}$ our method always leads to spreading of the density profile which costs almost zero free energy and it is interesting that a very recent study of Lutsko [20] based on a square gradient functional and parameterized profiles finds very similar behaviour for small Lennard Jones droplets. Although such behaviour is not fully understood we emphasize that following the genesis of the critical droplet is straightforward using our method.

In Fig 7 we plot the nucleation barrier height, the maximum of $\Delta\Omega$, for two inverse temperatures (a) $0.7\varepsilon_c^{-1}$ and (b) $0.8\varepsilon_c^{-1}$ for four different values of $A$ as a function of the degree of supersaturation. In the main figures we measure the latter in terms of the ordinate $x = (\mu_{spin} - \mu)/(\mu_{spin} - \mu_{bin})$, where $\mu_{spin}$ denotes the chemical potential at the spinodal. We believe that this is the best way to make comparisons of barrier heights between model fluids with different values of $A$; recall we are comparing at the same $\varepsilon^{-1}/\varepsilon_c^{-1}$ (essentially equivalent to comparing at the same $T/T_c$). The insets in Fig 7 plot the same data but now as function of the ratio of pressures $S^p = p(\mu)/p_{bin}$, where $p_{bin}$ is the saturated gas pressure at the particular value of $\varepsilon^{-1}$. The quantity $S^p$ is often used in nucleation studies – see [13] and references therein.

As $\mu \to \mu_{spin}$ the nucleation barrier must vanish as is clear in the plots. The interpretation of the results is that for a fixed $\varepsilon^{-1}/\varepsilon_c^{-1}$ and a given value of $x$, the nucleation barrier height decreases rapidly with increasing $A$.

Since the rate of nucleation decreases exponentially with $\beta \Delta\Omega$ this implies that the nucleation rate at a given supersaturation $x$ is much higher for the mermaid fluids than for the one-Yukawa model and increases with increasing $A$. Alternatively one can say that to achieve a given nucleation rate one must move closer to the spinodal as the amplitude $A$ is decreased. This is also evident from the plots in the insets. Note that $S_{spin}^p = p_{spin}/p_{bin}$ increases rapidly with increasing $A$. However, if one fixes the ratio $S^p$ at a particular value, say $S^p = 2.5$, then the height of the nucleation barrier increases rapidly with increasing $A$. Clearly the choice of ordinate to describe the degree of supersaturation is important in the interpretation of the results.

4.     **Concluding Remarks**

We have investigated the nucleation of liquid droplets for a model fluid described by the mermaid pair potential (1) using a simple DFT approach. We find that even for very small values of the amplitude $A$ the form of the droplet density profile and the height of the nucleation barrier are altered dramatically from the case $A=0$, i.e. the presence of a *very weak* long-ranged repulsive tail has a profound effect on the nucleation properties. Competing interactions influence not only the structure and thermodynamics of the supercritical fluid, effectively extending the critical region beyond that in a simple fluid [1], but also influence properties in the sub-critical, nucleation regime. In both cases one can regard the changes of behaviour observed at small $A$ as important precursors of the onset of inhomogeneous phases that occurs at higher values of this parameter. Our results should have repercussions for droplet nucleation rates in weakly charged, weakly screened colloidal systems. It is interesting to note that in a simulation study [21] of nucleation of the crystal from the fluid phase for a system of hard spheres with a single repulsive Yukawa tail , corresponding to setting $\varepsilon=0$ in (1), Auer and Frenkel found that as the strength of the repulsive tail was increased, nucleation barriers were decreased. This was attributed to the soft repulsion lowering the interfacial tension between the liquid and crystal phases [21, 22]. There is a similar effect in our present system: the long-range repulsion decreases the interfacial tension between the liquid and the gas phase.

The method used to implement the DFT for nucleation is different from that used in other DFT studies of droplets or bubbles. We believe it is very robust for finding the critical droplet; for all the state points we considered we were able to determine the nucleation barrier height. It does not suffer from the drawbacks of the original Oxtoby-Evans iteration scheme [10]. Requiring the magnitude of $\beta V_{eff}(r)$ to be everywhere small provides a valuable condition to check that one is indeed close to a saddle point in the grand potential and therefore to the critical droplet. There remain issues of interpretation of our results for values of $N_{ex}$ that do not correspond to the critical droplet. As remarked earlier, in these cases where the effective potential is non-zero, it is not clear whether the values of $\Delta\Omega$ we obtain are close to those one might obtain from other methods that purport to obtain the MFEP. Further study is required. We should also comment on the existence of two branches of $\Delta\Omega$ as a function of $N_{ex}$ corresponding to 'droplet' and 'flat' density profiles. In our numerical iteration we find that, for a given nucleation state point, starting from large $N_{ex}$ the solution is on a 'droplet' branch and that as $N_{ex}$ is decreased, $\Delta\Omega$ reaches a maximum. On further decreasing $N_{ex}$, the grand potential decreases before a discontinuous jump occurs to the 'flat' branch where $\Delta\Omega \approx 0$, corresponding to the excess particles being spread throughout the system. If one starts from small values of $N_{ex}$ the solution is on the 'flat' branch and increasing $N_{ex}$ eventually leads to a jump to the 'droplet' branch. The value of $N_{ex}$ at which the jump occurs increases with increasing system size $L$.

In Section 3 we discussed in detail the form of the droplet profiles for different $A$. The existence of long wavelength oscillations reflects the shape of the mermaid potential; competing interactions lead to the propensity towards clustering which is captured by the simple DFT. However, our DFT is a mean-field treatment, as is clear from the RPA form in (20), and one should ask how the oscillations in the droplet profile might be eroded by thermal fluctuations. Easier to analyse is a somewhat equivalent situation that arises at the planar gas-liquid interface discussed in [7]. For certain amplitudes $A$ and values of $\varepsilon^{-1}$ the RPA DFT predicts long wavelength oscillatory decay of the density profile into one or both of the coexisting fluid phases. In Figure 11 of [7] we give an example, for $A=0.082$, where both coexisting phases lie inside the upper Kirkwood line and there is oscillatory decay profile into both phases. For $A= 0.02$ (see the phase diagram in Figure 4 of [7]) there are also coexisting fluid phases, not too far below the critical point, which lie between two Kirkwood lines and where the DFT density profile exhibits oscillatory decay into both the gas and liquid bulk phases. The inverse decay length $\alpha_0$ and the wavelength $2\pi/\alpha_1$ of these oscillations are those obtained from the pole analysis that we described in Section 3. For the planar interface capillary wave fluctuations are expected to reduce the amplitude of the oscillations in the (intrinsic DFT interfacial profile) by a factor of exp $(-\alpha_1^2 \xi_{th}^2/2)$ where $\xi_{th}$ is the thermal roughness of the interface. The Gaussian convolution on which this result is based – see [19] and references therein – is usually applied to simple fluids where the wavelength of oscillations is much smaller, about $\sigma$. Thus for a given thermal roughness one might expect the dampening factor of the long wavelength oscillations we encounter for the present models with $A>0$ to be considerably smaller than for the simple fluids. Of course, fluctuation effects in metastable droplets are much more subtle-see e.g. [12] and an interesting, more general, discussion on the role of fluctuations in DFT [23].

One should also enquire about the limitations of our simple mean-field DFT for describing the bulk phase behaviour and fluid structure in the vicinity of a $\lambda$ line. When a mean-field theory predicts that the uniform fluid becomes unstable w.r.t. periodic density fluctuations this usually indicates that a transition to a modulated fluid (cluster) phase should have occurred earlier in an 'exact' treatment that incorporates all relevant fluctuations. Note that SCOZA does not exhibit a $\lambda$ line; instead the theory has regions of the phase diagram where the SCOZA equations cannot be solved [7]. By imposing particular forms for density modulations one can use the present RPA DFT to investigate the onset of transitions to modulated phases [8, 24] - see also [25] describing a DFT study of microphase separation in a two-dimensional fluid. In the present study we deliberately avoid the close vicinity of a $\lambda$ line and for $A=0.06$, where this line lies on top of the spinodal, we deliberately perform our nucleation calculations at reduced 'temperatures' well below the critical point where gas-liquid phase separation should not be affected by the possible existence of a modulated phase [8]. For smaller values of $A$, modulated phases do not appear.

Finally we remark that gas (bubble) nucleation has been a topic of some debate in the recent literature [14, 15, 26, 27]. The method we implement here is readily applicable to the calculation of the critical bubble in a superheated liquid. The same caveats that we mentioned for non-critical droplets apply to non-critical bubbles. For the mermaid potential, $A>0$, state points corresponding to the metastable bulk liquid will generally lie between the two Kirkwood lines. This implies that the tail of the density profile of the bubble will exhibit long wavelength oscillations, i.e., $\rho_b - \rho(r) \sim B\exp(-\alpha_0 r)\cos(\alpha_1 r - \theta)/r$, $r \to \infty$, where the parameters are now determined by the density $\rho_b$ and 'temperature' of the metastable bulk *liquid*. It is likely that the amplitude of oscillations is larger on decay into the liquid than into the gas.

**Acknowledgements**

It is a pleasure to dedicate this contribution to Luciano Reatto-a hugely influential figure in the theory of liquids. He and Davide Pini inspired both of us to investigate the structural properties and phase behaviour of mermaid models. We are grateful for their insight and for guiding us to many of the problems, and the solutions, that we addressed in recent years. AJA acknowledges the support of RCUK.

**Figures**

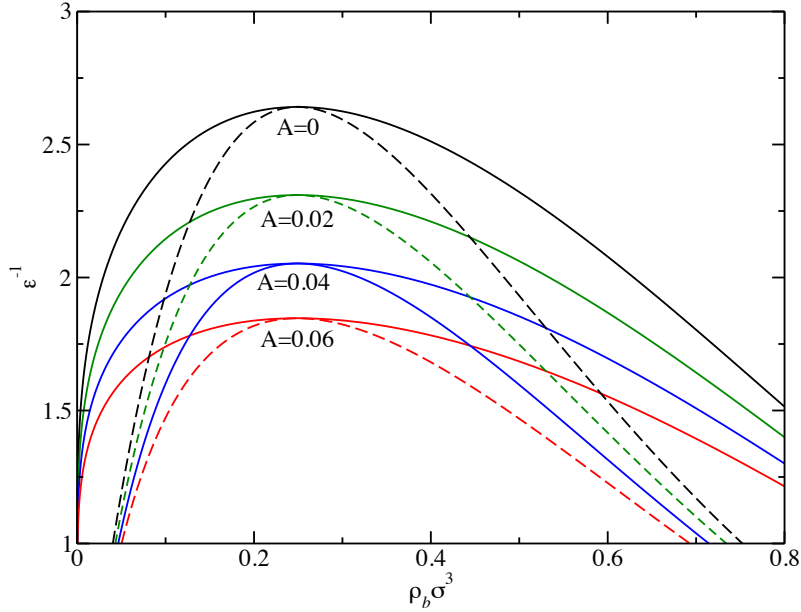

Figure 1. Phase diagram for various values of $A$, the amplitude of the repulsive Yukawa tail, in the 'temperature' $\varepsilon^{-1}$ – density plane for $Z_1 = 1.0$ and $Z_2 = 0.5$. The solid lines denote the gas-liquid binodals and the dashed lines the spinodals. Note that for $A = 0.06$ there is a $\lambda$ line (not shown) that is almost indistinguishable from the spinodal – see text.

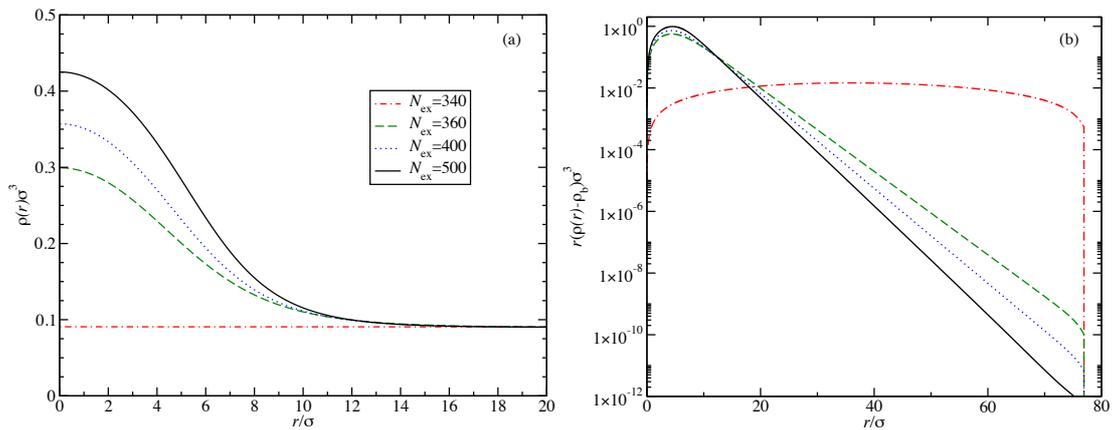

Figure 2. (a) Droplet density profiles, plotted as a function of the distance from the centre, for various values of $N_{ex}$, the excess number of particles. $A = 0$, $\varepsilon^{-1} = 0.8\varepsilon_c^{-1}$ and the (metastable) bulk gas density is $\rho_b\sigma^3 = 0.09$. The droplet with $N_{ex} = 400$ corresponds roughly to the critical droplet – see Figure 6. (b) The same profiles plotted as $r(\rho(r) - \rho_b)$ on a logarithmic scale, confirming the exponential decay predicted by (23). For the smallest

droplet, $N_{ex}$ = 340 (dot-dash line), the excess particles are spread throughout the system. Note that the system size $L = 77\sigma$ for all the calculations.

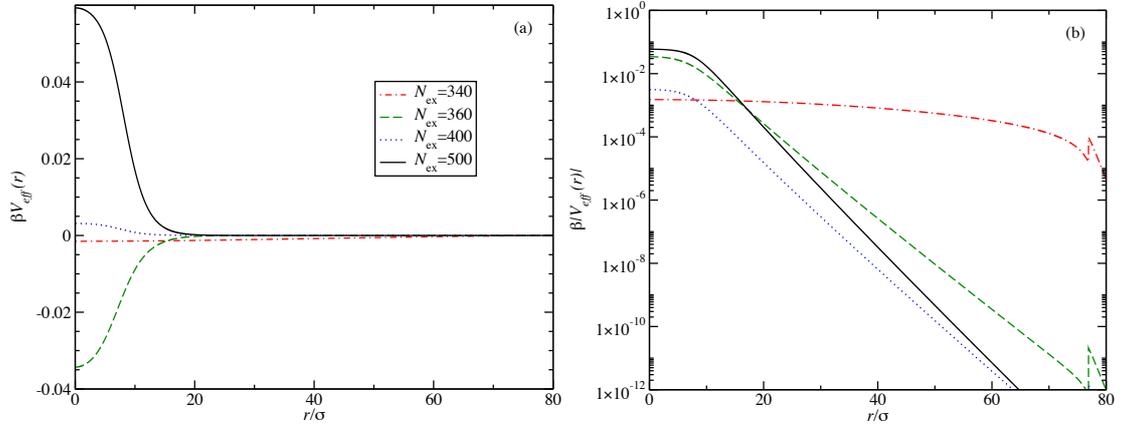

Figure 3. (a) The effective external potential $V_{eff}(r)$ that generates the density profiles in Figure 2. The potential corresponding to the near critical droplet $N_{ex}$ = 400 is much weaker than those for $N_{ex}$ = 500 and 360. (b) $\beta|V_{eff}(r)|$ plotted on a logarithmic scale.

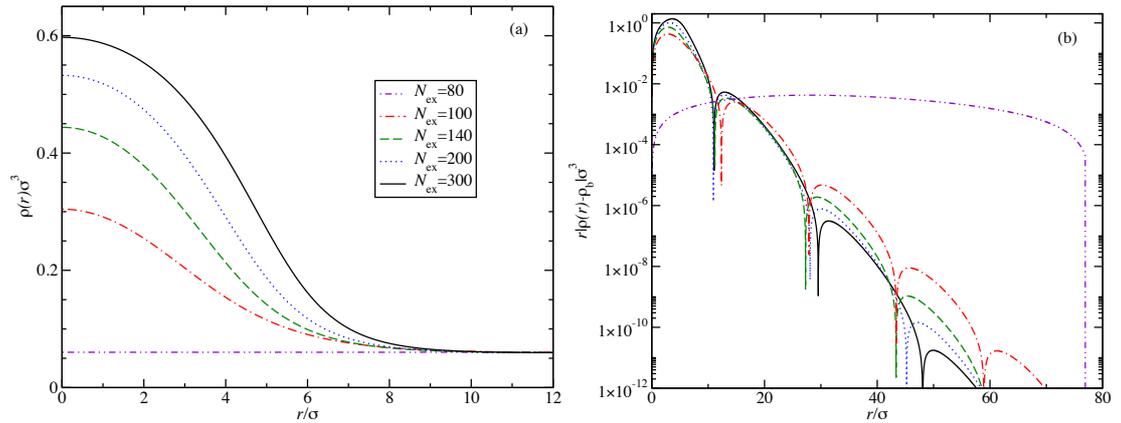

Figure 4. (a) As in Figure 2, except the density profiles are for the case with $A = 0.06$, $\varepsilon^{-1} = 0.8\varepsilon_c^{-1}$ and $\rho_b\sigma^3 = 0.06$. The droplet with $N_{ex}$ = 140 is near critical – see Figure 6. (b) The same profiles plotted as $r|\rho(r) - \rho_b|$ on a logarithmic scale illustrating the exponentially damped, long wavelength oscillatory decay associated with a propensity towards clustering in this fluid. For the near critical droplet, $N_{ex}$ = 140 (dashed line), the wavelength of the oscillations is about $32\sigma$; the wavelength and decay length are predicted correctly by the pole analysis leading to (24).

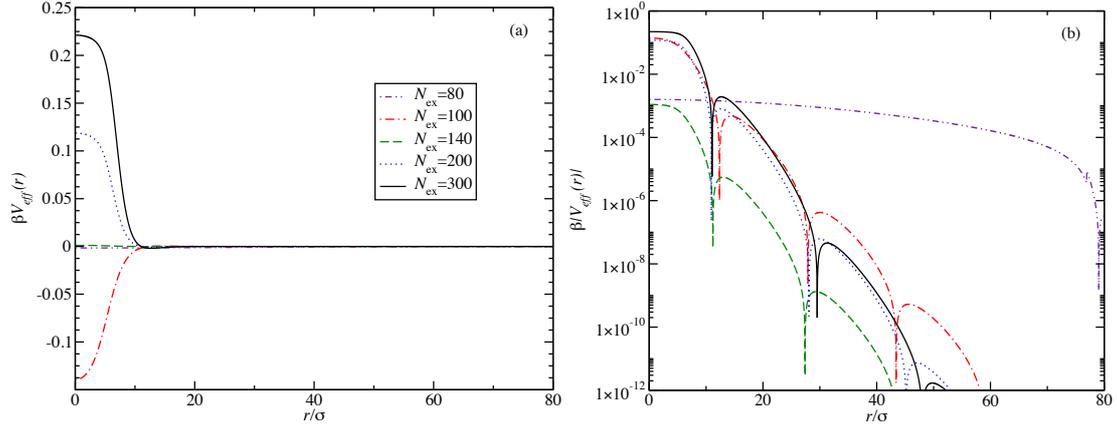

Figure 5. (a) The effective potentials $V_{eff}(r)$ that generate the density profiles in Figure 4. The potential for the near-critical droplet $N_{ex} = 140$ is much weaker than those for the other droplets. (b) $\beta|V_{eff}(r)|$ plotted on a logarithmic scale. The wavelength of the oscillations is the same as that of the corresponding profile in Figure 4.

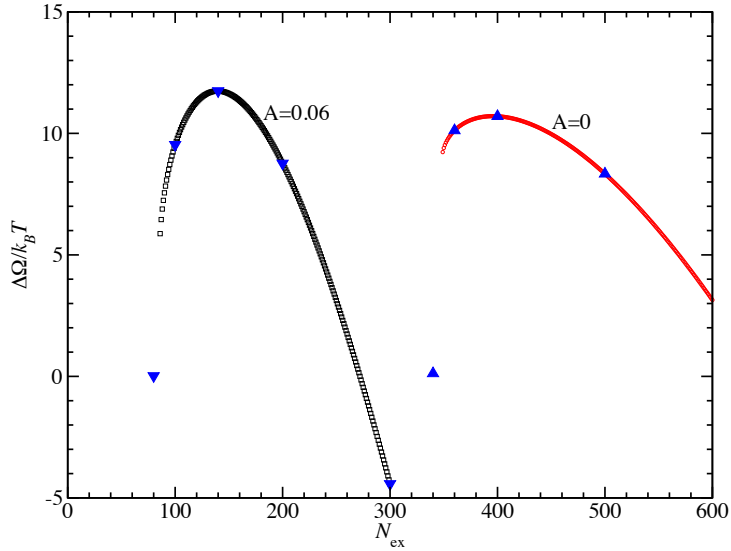

Figure 6. The excess grand potential $\Delta\Omega$ as a function of the excess number of particles $N_{ex}$ in the droplet for i) $A=0$ and $\rho_b\sigma^3 = 0.09$ and ii) $A=0.06$ and $\rho_b\sigma^3 = 0.06$. In both cases $\varepsilon^{-1} = 0.8\varepsilon_c^{-1}$. For the points marked with a triangle the corresponding droplet density profiles are displayed in Figures 2 and 4. Note that for the small droplets with $N_{ex} = 340$ ($A=0$) and $N_{ex} = 80$ ($A=0.06$) the excess of particles is spread throughout the system and $\Delta\Omega$ is zero.

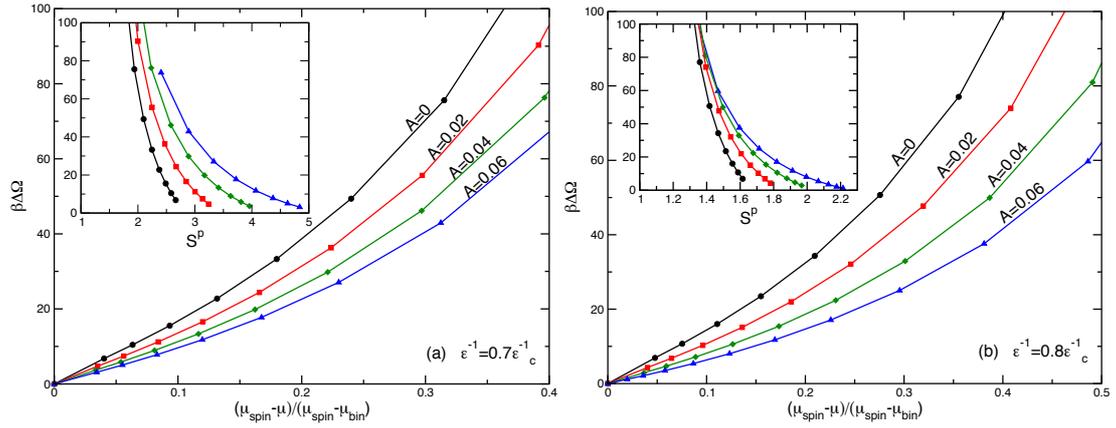

Figure 7. The nucleation free energy barrier height as a function of (scaled) chemical potential $\mu$. $\mu_{spin}$ is the value at the spinodal and $\mu_{bin}$ at the binodal. (a) reduced 'temperature' $\varepsilon^{-1} = 0.7\varepsilon_c^{-1}$ and (b) $\varepsilon^{-1} = 0.8\varepsilon_c^{-1}$. For a fixed value of the ordinate increasing the amplitude $A$ of the repulsive Yukawa tail decreases the barrier height and thereby increases significantly the nucleation rate. The insets show the same data but now plotted as a function of the ratio of pressures $S^p = p(\mu)/p_{bin}$. It is important to note that the value of $S^p$ at the spinodal, where the barrier height vanishes, increases rapidly with $A$.